\documentclass[prd,aps,showpacs,tightenlines,nofootinbib,preprint,eqsecnum]
{revtex4}

\usepackage{epsf}
\usepackage{graphics}
\usepackage{amssymb}
\usepackage{amsmath}

\newcommand{\bear}{\begin{array}}  \newcommand{\eear}{\end{array}}
\newcommand{\bea}{\begin{eqnarray}}  \newcommand{\eea}{\end{eqnarray}}
\newcommand{\beq}{\begin{equation}}  \newcommand{\eeq}{\end{equation}}
\newcommand{\bef}{\begin{figure}}  \newcommand{\eef}{\end{figure}}
\newcommand{\bec}{\begin{center}}  \newcommand{\eec}{\end{center}}

\newcommand{\bib}{\bibitem}

\newcommand{\Eqn}[1]{&\hspace{-0.2em}#1\hspace{-0.2em}&}

\def\APJ#1#2#3{Astrophys. J. {\bf #1}, #2 (19#3)}
\def\APJJ#1#2#3{Astrophys. J. {\bf #1}, #2 (20#3)}

\def\ARNPS#1#2#3{Ann. Rev. Nucl. Part. Sci. {\bf#1}, #2 (19#3)}

\def\IJMPDD#1#2#3{Int. J. Mod. Phys. D {\bf #1}, #2 (20#3)}

\def\NPB#1#2#3{Nucl. Phys. {\bf B#1}, #2 (19#3)}

\def\PLB#1#2#3{Phys. Lett. B {\bf #1}, #2 (19#3)}

\def\PL#1#2#3{Phys. Lett. {\bf #1}, #2 (19#3)}
\def\PLBold#1#2#3{Phys. Lett. {\bf#1B}, #2 (19#3)}

\def\PRD#1#2#3{Phys. Rev. D {\bf #1}, #2 (19#3)}
\def\PRDD#1#2#3{Phys. Rev. D {\bf #1}, #2 (20#3)}

\def\PRL#1#2#3{Phys. Rev. Lett. {\bf#1}, #2 (19#3)}

\def\PRT#1#2#3{Phys. Rep. {\bf#1}, #2 (19#3)}
\def\PRTT#1#2#3{Phys. Rep. {\bf#1}, #2 (20#3)}
\def\PTP#1#2#3{Prog. Theor. Phys. {\bf #1}, #2 (19#3)}
\def\PTPP#1#2#3{Prog. Theor. Phys. {\bf #1}, #2 (20#3)}

\def\RMP#1#2#3{Rev. Mod. Phys. {\bf #1}, #2 (19#3)}
\def\RMPP#1#2#3{Rev. Mod. Phys. {\bf #1}, #2 (20#3)}

\def\Vec#1{\mbox{\boldmath $#1$}}
\def\Vecs#1{\mbox{\boldmath\tiny $#1$}}

\begin{document}

\title{Baryon asymmetry from hypermagnetic helicity 
in dilaton hypercharge electromagnetism 
      }

\author{Kazuharu Bamba\footnote{
Research Fellow of the Japan Society for the Promotion of Science}
\footnote{E-mail: bamba@yukawa.kyoto-u.ac.jp}
}  
\affiliation{
Yukawa Institute for Theoretical Physics, 
Kyoto University, 
Kyoto 606-8502, 
Japan}


\begin{abstract}

The generation of the baryon asymmetry of the Universe (BAU) from 
the hypermagnetic helicity, 
the physical interpretation of which is 
given in terms of hypermagnetic knots, 
is studied in inflationary cosmology, taking into account the 
breaking of the conformal invariance of hypercharge electromagnetic fields 
through both a coupling with the dilaton and that with a pseudoscalar field.  
It is shown that if the electroweak phase transition (EWPT) is strongly first 
order and the present amplitude of the generated magnetic fields on 
the horizon scale is sufficiently large, 
a baryon asymmetry with a sufficient magnitude to 
account for the observed baryon to entropy ratio can be generated.   

\end{abstract}


\hspace{13.0cm} YITP-06-47

\maketitle

\section{Introduction}

It is observationally known that 
there exists a net excess of baryons over antibaryons in 
the Universe.  
In order to generate the baryon asymmetry of the Universe (BAU) 
from a state in which baryons and antibaryons have same abundances, 
the Sakharov's three conditions \cite{Sakharov1} must be satisfied:  
(1) baryon number nonconservation, (2) $C$ and $CP$ violation, 
and (3) a departure from thermal equilibrium.  
Many cosmological scenarios in which the above three conditions 
could be satisfied have been proposed 
(for reviews of the BAU, see 
Refs.~\cite{Dolgov92, DK04}).  
The origin of the BAU, 
however, is not well established yet.  

It has been pointed out 
\cite{JS97, GS98-1, GS98-2, Thompson98, Vachaspati94} that 
hypercharge electromagnetic fields could play a significant role 
in the electroweak (EW) scenario 
\cite{Cohen93, Rubakov96, Funakubo96, Trodden99} for baryogenesis.  
Giovannini and Shaposhnikov \cite{GS98-1, GS98-2} have shown that 
the Chern-Simons number stored 
in the hypercharge electromagnetic 
fields, i.e., the hypermagnetic helicity, is converted into 
fermions at the electroweak phase transition (EWPT) owing to 
the Abelian anomaly \cite{Kuzmin85}, 
and at the same time 
the hypermagnetic fields are replaced by the ordinary magnetic fields, 
which survive after the EWPT.  
The hypermagnetic helicity corresponds to a topologically non-trivial 
hypercharge configurations.  
Topologically non-trivial refers here to the topology of the magnetic 
flux lines.  
The physical interpretation of these hypercharge configurations is 
given in terms of hypermagnetic knots. 
The generated fermions will not 
be destroyed by sphaleron processes \cite{Sphaleron} if the EWPT is 
strongly first order.  This condition could be met 
in the extensions of the minimal standard model (MSM) \cite{Funakubo05}.  

The most natural origin of large-scale hypermagnetic fields before 
the EWPT is hypercharge electromagnetic quantum fluctuations generated 
in the inflationary stage 
\cite{Turner} (for reviews of inflation, see 
Refs.~\cite{Linde1, Kolb}).  
This is because inflation naturally produces effects on very large scales, 
larger than Hubble horizon, starting from microphysical processes 
operating on a causally connected volume.  
If the conformal invariance of the Maxwell theory is broken by some 
mechanism in the inflationary stage,  
hypercharge electromagnetic quantum fluctuations could be generated 
even in the Friedmann-Robertson-Walker (FRW) spacetime, which 
is conformally flat.  
Hence several breaking mechanisms have been proposed in 
Refs.~\cite{Turner, Ratra, Scalar, Garretson92, Field00, 
Dolgov93, Bamba1, Bamba2} 
(for other mechanisms, see reviews of the origin of cosmic magnetic fields 
\cite{Grasso01, Dolgov01, Widrow02, Giovannini04, Semikoz05} 
and references therein).   
In particular, it follows from indications 
in higher-dimensional theories including string theory that 
there can exist the dilaton field coupled to the hypercharge 
electromagnetic fields.  This coupling also breaks the conformal invariance of 
the Maxwell theory \cite{Ratra, Scalar, Bamba1, Bamba2} 
and hence the large-scale hypercharge magnetic fields could be generated.  

Furthermore, it has been noticed \cite{BO99, Giovannini00} that 
when a pseudoscalar field $\phi$ with an axion-like coupling to 
the $U(1)_Y$ hypercharge field strength $Y_{\mu\nu}$ 
in the form $\phi Y_{\mu\nu}\tilde{Y}^{\mu\nu}$, where 
$\tilde{Y}^{\mu\nu}$ is the dual of $Y_{\mu\nu}$ and 
$Y_{\mu\nu}\tilde{Y}^{\mu\nu}$ 
corresponds to the hypercharge topological number density, 
coherently rolls or oscillates before the EWPT, 
the Sakharov's three conditions can be satisfied.  
In this case, the motion of the pseudoscalar field generates 
a time-dependent hypercharge topological number condensate 
which violates fermion number conservation through the Abelian 
anomalous coupling \cite{Kuzmin85}, and realizes a departure from 
equilibrium.  Moreover, $C$ symmetry is violated by the chiral coupling 
of the hypercharge gauge boson to fermions.  Furthermore, since 
the hypercharge topological number is odd under $CP$, 
$CP$ symmetry is spontaneously broken if the hypercharge topological number 
is coupled to a field with a time-dependent expectation value.  
This mechanism is able to generate a net Chern-Simons number which can 
survive until the EWPT and then be converted into a baryon asymmetry.  
Pseudoscalar fields with the above axion-like coupling appear in 
several possible extensions of the standard model \cite{BO01}.  
Incidentally, the generation of large-scale magnetic fields owing to 
the breaking of the conformal invariance of the Maxwell theory through such an 
axion-like coupling between a pseudoscalar field and 
electromagnetic fields have been considered 
in Refs.~\cite{Turner, Garretson92, Field00}.  
Moreover, 
baryogenesis due to the above coupling has been discussed in 
Ref.~\cite{Guendelman92}.  
Furthermore, the generation of the magnetic helicity 
owing to the above coupling has been considered in Ref.~\cite{Campanelli05}.   

In the present paper,  
in addition to the existence of the dilaton coupled to hypercharge 
electromagnetic fields, 
we assume the existence of a pseudoscalar field with an 
axion-like coupling to hypercharge electromagnetic fields, 
and consider the generation of the BAU in slow-roll exponential 
inflation models.  
In particular, we consider the generation of the 
Chern-Simons number, i.e., the hypermagnetic helicity,  
through the coupling between 
a pseudoscalar field and hypercharge electromagnetic fields 
in the inflationary stage.  
The generated Chern-Simons number is converted into 
fermions at the EWPT owing to the Abelian anomaly, 
and the generated fermions can survive after the EWPT 
if the EWPT is strongly first order.   
The reason why we consider the generation of the 
Chern-Simons number during inflation is as follows:  
In this scenario, 
the BAU is induced by the helicity of 
the hypermagnetic fields.  Hence 
the large-scale hypermagnetic fields whose present scale is 
larger than the present horizon scale have to be generated 
in order that the resultant BAU could be 
homogeneous over the present horizon.

This paper is organized as follows.  
In Sec.\ II we describe our model and derive equations of motion 
from its action.  
In Sec.\ III we consider the evolution of the $U(1)_Y$ gauge field 
and investigate the generated Chern-Simons number density, 
and then estimate the resultant baryon asymmetry.  
Finally, Sec.\ IV is devoted to a conclusion.  

We use units in which $k_\mathrm{B} = c = \hbar = 1$ and denote the 
gravitational constant $8 \pi G$ by ${\kappa}^2$ so that 
${\kappa}^2 \equiv 8\pi/{M_{\mathrm{Pl}}}^2$ where 
$M_{\mathrm{Pl}} = G^{-1/2} = 1.2 \times 10^{19}$GeV is the Planck mass.  
Moreover, in terms of electromagnetism we adopt Heaviside-Lorentz units.  
Throughout the present paper, 
the subscripts `1', `R', and `0' represent the quantities at  
the time $t_1$ when a given mode of the $U(1)_Y$ gauge field 
first crosses outside the horizon during inflation, 
the end of inflation (namely, the instantaneous reheating stage) 
$t_\mathrm{R}$, and the present time $t_0$, respectively.

\section{MODEL}

\subsection{Action}

We introduce the dilaton field $\Phi$ and 
a pseudoscalar field $\phi$.   
Moreover, we introduce the coupling of the dilaton to hypercharge 
electromagnetic fields and that of the pseudoscalar to those fields.  
Our model action is the following:  
\begin{eqnarray}
S 
\Eqn{=} 
\int d^{4}x \sqrt{-g} 
\left[ \hspace{1mm}
{\mathcal{L}}_{\mathrm{inflaton}}   
+
{\mathcal{L}}_{\mathrm{dilaton}}  
+ 
{\mathcal{L}}_{\mathrm{ps}}  
+
{\mathcal{L}}_{\mathrm{HEM}} 
\hspace{1mm} \right],  
\label{eq:2.1}  \\[3mm]
{\mathcal{L}}_{\mathrm{inflaton}}   
\Eqn{=} 
-\frac{1}{2}g^{\mu\nu}{\partial}_{\mu}{\varphi}{\partial}_{\nu}{\varphi} 
- U[\varphi],  
\label{eq:2.2} \\[3mm]
{\mathcal{L}}_{\mathrm{dilaton}}  
\Eqn{=}
-\frac{1}{2}g^{\mu\nu}{\partial}_{\mu}{\Phi}{\partial}_{\nu}{\Phi} 
- V[\Phi],  
\label{eq:2.3} 
\end{eqnarray} 
\begin{eqnarray}
{\mathcal{L}}_{\mathrm{ps}}  
\Eqn{=}
-\frac{1}{2}g^{\mu\nu}{\partial}_{\mu}{\phi}{\partial}_{\nu}{\phi} 
- W[\phi],  
\label{eq:2.4} \\[3mm]
{\mathcal{L}}_{\mathrm{HEM}}  \Eqn{=} 
-\frac{1}{4} 
f(\Phi) \left(
Y_{\mu\nu}Y^{\mu\nu} + g_{\mathrm{ps}} \frac{\phi}{M} 
 Y_{\mu\nu}\tilde{Y}^{\mu\nu} \right),  
\label{eq:2.5}  \\[3mm]  
f(\Phi) \Eqn{=} \exp \left(-\lambda \kappa \Phi \right), 
\label{eq:2.6}  
\\[3mm] 
V[\Phi] \Eqn{=} \bar{V} \exp \left( -\tilde{\lambda} \kappa \Phi \right), 
\label{eq:2.7}  \\[3mm] 
W[\phi] \Eqn{=} \frac{1}{2} m^2 \phi^2, 
\label{eq:2.8} 
\end{eqnarray}  
where $g$ is the determinant of the metric tensor $g_{\mu\nu}$, 
$U[\varphi]$, $V[\Phi]$, and $W[\phi]$ are the inflaton, the dilaton, and 
the pseudoscalar potentials, respectively, 
$\bar{V}$ is a constant, $m$ is the mass of the pseudoscalar field, 
and 
$f$ is the coupling between the dilaton and hypercharge 
electromagnetic fields with 
$\lambda$\footnote{
The sign of $\lambda$ in the present paper is opposite to that in 
Refs.~\cite{Bamba1, Bamba2}
} 
and $\tilde{\lambda} \hspace{0.5mm} 
(\hspace{0.5mm} > 0 \hspace{0.5mm})$ being 
dimensionless constants.  
Moreover, 
$g_{\mathrm{ps}} = \bar{g}_{\mathrm{ps}} \alpha^{\prime}/(2\pi)$, 
where $\bar{g}_{\mathrm{ps}}$ is a numerical factor and 
$\alpha^{\prime} = g^{\prime \hspace{0.3mm} 2}/(4\pi)$.  Here, $g^{\prime}$ is 
the $U(1)_Y$ gauge coupling constant.  
$M$ denotes a mass scale. 
Furthermore, the $U(1)_Y$ hypercharge field strength is given by 
$Y_{\mu\nu} = \nabla_{\mu} Y_{\nu} - \nabla_{\nu} Y_{\mu}$, where 
$\tilde{Y}^{\mu\nu} = (1/2) \epsilon^{\mu\nu\rho\sigma} Y_{\rho\sigma}$.  
Here, $Y_{\mu}$ is the $U(1)_Y$ gauge field, $\nabla_{\mu}$ denotes 
the covariant derivative, and $\epsilon^{\mu\nu\rho\sigma}$ is the 
Levi-Civita tensor.  
Moreover, we note that 
covariant derivatives for the antisymmetric tensor $Y_{\mu\nu}$ in the 
metric (2.13) as is shown in the next subsection 
are simple derivatives, $\nabla_{\mu}=\partial_{\mu}$, 
as 
it appears later in the equations of motion (2.12) and (2.22) based on 
the hypercharge electromagnetic part of our model 
Lagrangian in Eq.\ (\ref{eq:2.5}).  

\subsection{Equations of motion}

From the above action in Eq.\ (\ref{eq:2.1}), the equations of motion for 
the inflaton, the dilaton, the pseudoscalar field, and hypercharge 
electromagnetic fields can be derived 
as follows:  
\begin{eqnarray}
-\frac{1}{\sqrt{-g}}{\partial}_{\mu} 
\left( \sqrt{-g}g^{\mu\nu}{\partial}_{\nu} \varphi \right)
+ \frac{d U[\varphi]}{d \varphi} \Eqn{=} 0, 
\label{eq:2.9}  \\[3mm] 
-\frac{1}{\sqrt{-g}}{\partial}_{\mu} 
\left( \sqrt{-g}g^{\mu\nu}{\partial}_{\nu} \Phi \right)
+ \frac{dV[\Phi]}{d\Phi} \Eqn{=} 
-\frac{1}{4} \frac{d f(\Phi)}{d \Phi}
\left( Y_{\mu\nu}Y^{\mu\nu} + g_{\mathrm{ps}} \frac{\phi}{M} 
Y_{\mu\nu} \tilde{Y}^{\mu\nu} \right),  
\label{eq:2.10}  \\[3mm] 
-\frac{1}{\sqrt{-g}}{\partial}_{\mu} 
\left( \sqrt{-g}g^{\mu\nu}{\partial}_{\nu} \phi \right)
+ \frac{d W[\phi]}{d \phi} \Eqn{=} 
-\frac{1}{4} \frac{g_{\mathrm{ps}}}{M} 
f(\Phi) Y_{\mu\nu}\tilde{Y}^{\mu\nu},  
\label{eq:2.11}  \\[3mm] 
\frac{1}{\sqrt{-g}}{\partial}_{\mu} 
\left( \sqrt{-g} f(\Phi) Y^{\mu\nu} \right) 
\Eqn{=}  - \frac{g_{\mathrm{ps}}}{M} 
{\partial}_{\mu} \left[ f(\Phi) \phi \right] 
\tilde{Y}^{\mu\nu}.  
\label{eq:2.12} 
\end{eqnarray} 

We now assume the spatially flat 
Friedmann-Robertson-Walker (FRW) space-time with the metric 
\begin{eqnarray}
{ds}^2 = g_{\mu\nu}dx^{\mu}dx^{\nu} 
\Eqn{=}   -{dt}^2 + a^2(t)d{\Vec{x}}^2 \nonumber \\[3mm]
\Eqn{=}   a^2(\eta) ( -{d \eta}^2 + d{\Vec{x}}^2 ), 
\label{eq:2.13}
\end{eqnarray} 
where $a$ is the scale factor, and $\eta$ is the conformal time.  

Since we are interested in the specific case in which 
the background space-time is inflating, we assume that the spatial 
derivatives of $\varphi$, $\phi$, and $\Phi$ are negligible 
compared to the other terms 
(if this is not the case at the beginning of inflation, any spatial 
inhomogeneities will quickly be inflated away and this assumption will 
quickly become very accurate).  
Hence the equations of motion for the background homogeneous scalar fields 
read 
\begin{eqnarray} 
\ddot{\varphi} + 3H\dot{\varphi} + \frac{dU[\varphi]}{d\varphi} \Eqn{=} 0,
\label{eq:2.14}  
\end{eqnarray} 
\begin{eqnarray} 
\ddot{\Phi} + 3H\dot{\Phi} + \frac{dV[\Phi]}{d\Phi} \Eqn{=} 0,
\label{eq:2.15}  \\[3mm]
\ddot{\phi} + 3H\dot{\phi} + \frac{dW[\phi]}{d\phi} \Eqn{=} 0, 
\label{eq:2.16}
\end{eqnarray} 
together with the background Friedmann equation
\begin{eqnarray}
H^2 \Eqn{=} \left( \frac{\dot{a}}{a} \right)^2 = \frac{{\kappa}^2}{3}
 \left( {\rho}_{\varphi} + {\rho}_{\Phi} + {\rho}_{\phi} \right),  
\label{eq:2.17} \\[3mm]
{\rho}_{\varphi} \Eqn{=} \frac{1}{2}{\dot{\varphi}}^2 + U[\varphi], 
\label{eq:2.18} \\[3mm]
{\rho}_{\Phi} \Eqn{=} \frac{1}{2}{\dot{\Phi}}^2 + V[\Phi], 
\label{eq:2.19} \\[3mm]
{\rho}_{\phi} \Eqn{=} \frac{1}{2}{\dot{\phi}}^2 + W[\phi], 
\label{eq:2.20} 
\end{eqnarray}
where a dot denotes a time derivative.  
Here, ${\rho}_{\varphi}$, ${\rho}_{\Phi}$, and ${\rho}_{\phi}$
are the energy densities of the inflaton, the dilaton, and the 
pseudoscalar field, respectively.  
We here consider the case in which slow-roll exponential inflation 
is driven by the potential energy of the inflaton 
and during inflation the energy densities of the dilaton and 
the pseudoscalar field are much smaller than that of the inflaton, 
${\rho}_{\varphi} \gg {\rho}_{\Phi}$,  ${\rho}_{\varphi} \gg {\rho}_{\phi}$.  
Hence, during inflation $H$ reads 
\begin{eqnarray}
H^2 \approx \frac{{\kappa}^2}{3} {\rho}_{\varphi} \equiv {H_{\mathrm{inf}}}^2,
\label{eq:2.21}
\end{eqnarray} 
where $H_{\mathrm{inf}}$ is the Hubble constant in the inflationary stage.  

We consider the evolution of the $U(1)_Y$ gauge field in this background.  
Its equation of motion in the Coulomb gauge, 
$Y_0(t,\Vec{x}) = 0$ and ${\partial}_j Y^j (t,\Vec{x}) =0$, becomes 
\begin{eqnarray}
\ddot{Y_i}(t,\Vec{x}) 
+ \left( H + \frac{\dot{f}}{f} 
\right) \dot{Y_i}(t,\Vec{x}) 
- \frac{1}{a^2}{\partial}_j {\partial}_j Y_i(t,\Vec{x}) 
- \frac{g_{\mathrm{ps}}}{M} \frac{1}{af} \frac{d \left(f \phi \right)}{d t} 
\epsilon^{ijk} {\partial}_j Y_k(t,\Vec{x}) = 0. 
\label{eq:2.22}
\end{eqnarray}

\section{Generation of the BAU}

In this section, we consider the generation of the BAU.  
First, we consider 
the evolution of the dilaton, that of the pseudoscalar field, and 
that of the $U(1)_Y$ gauge field.  Next, we investigate the generation of the 
Chern-Simons number density in the inflationary stage, and then estimate 
the ratio of the baryonic number density to the entropy density.  

\subsection{Evolution of the dilaton and that of the pseudoscalar field}

In this subsection, 
we consider the evolution of the dilaton and the pseudoscalar field.  
Here we consider the case in which slow-roll exponential inflation is 
realized and the scale factor $a(t)$ is given by 
\begin{eqnarray}
a(t) = a_1 \exp \left[ \hspace{0.5mm} H_{\mathrm{inf}}(t-t_1) \right], 
\label{eq:3.1}
\end{eqnarray}
where $a_1$ is the scale factor at the time $t_1$ when a given 
comoving wavelength $2\pi/k$ of the $U(1)_Y$ gauge field 
first crosses outside the horizon during 
inflation, $k/(a_1 H_{\mathrm{inf}}) = 1$.  

First, we investigate the evolution of the dilaton.  
We consider the case in which 
we can apply slow-roll approximation to the dilaton, that is, 
\begin{eqnarray}               
  \left| \frac{\ddot{\Phi}}{H_{\mathrm{inf}}\dot{\Phi}} \right| \ll 1,  
\label{eq:3.2}
\end{eqnarray}
and then Eq.\ (\ref{eq:2.15}) is reduced to 
\begin{eqnarray}   
 3H_{\mathrm{inf}} \dot{\Phi} +  \frac{dV[\Phi]}{d\Phi} = 0. 
\label{eq:3.3}
\end{eqnarray}
The solution of this equation is given by 
\begin{eqnarray}                             
\Phi \Eqn{=} \frac{1}{\tilde{\lambda}\kappa}
\ln \left[ 
\tilde{\lambda}^2 w H_{\mathrm{inf}} \left( t-t_\mathrm{R} \right)  
+ \exp \left( \tilde{\lambda} \kappa {\Phi}_\mathrm{R} \right)
\right], 
\label{eq:3.4}  \\[3mm]
w \Eqn{\equiv} \frac{\bar{V}}{3 H_{\mathrm{inf}}^2/ \kappa^2} 
\approx \frac{\bar{V}}{\rho_{\varphi}}, 
\label{eq:3.5}
\end{eqnarray} 
where $\Phi_{\mathrm{R}}$ is 
the dilaton field amplitude at the end of inflation.   
Here we consider the case in which after inflation, the dilaton is finally 
stabilized when it feels other contributions to its potential, e.g., 
from gaugino condensation \cite{GC} that generates a potential minimum 
\cite{Barreiro, Seto}.  
As it reaches there, the dilaton starts oscillation and 
finally decays into radiation at $t=t_\mathrm{R}$.  
Hence we assume that the potential minimum is generated at 
$\Phi = \Phi_{\mathrm{R}} = 0$, so that 
the coupling $f$ between the dilaton and hypercharge electromagnetic fields is 
set to unity and thus the standard Maxwell theory is recovered.  
Moreover, in deriving the second approximate equality in Eq.\ (\ref{eq:3.5}), 
we have used Eq.\ (\ref{eq:2.21}).  
Since we have ${\rho}_{\varphi} \gg {\rho}_{\Phi}$ by assumption, $w \ll 1$.  

It follows from Eq.\ (\ref{eq:3.4}) that 
the slow-roll condition to the dilaton, Eq.\ (\ref{eq:3.2}), 
is equivalent to the following relation:  
\begin{eqnarray}
\left| \frac{\ddot{\Phi}}{H_{\mathrm{inf}}\dot{\Phi}} \right| \ll 1  
\hspace{2mm} \Longleftrightarrow \hspace{2mm} 
\tilde{\lambda}^2 \frac{V[\Phi]}{{\rho}_{\varphi}} \ll 1.
\label{eq:3.6}
\end{eqnarray}
In deriving this relation, we have used Eq.\ (\ref{eq:2.21}).  
If we assume that ${\tilde{\lambda}} \sim \mathcal{O}(1)$, 
the second relation in (\ref{eq:3.6}), 
$\tilde{\lambda}^2 V[\Phi]/{\rho}_{\varphi} \ll 1$, 
is satisfied during inflation 
because ${\rho}_{\varphi} \gg {\rho}_{\Phi}$.  

Next, we consider the evolution of the pseudoscalar field.  
The solution of Eq.\ (\ref{eq:2.16}) with Eq.\ (\ref{eq:2.8}) is 
given by \cite{Garretson92} 
\begin{eqnarray} 
\phi = \phi_1 \exp \left\{ 
\frac{3}{2} \left[-1 \pm \sqrt{1- \left( \frac{2m}{3H_{\mathrm{inf}}} 
\right)^2} \right] H_{\mathrm{inf}} \left( t-t_1 \right)
\right\}.
\label{eq:3.7} 
\end{eqnarray}
For $m \gg H_{\mathrm{inf}}$, 
we find that the approximate solution 
of Eq.\ (\ref{eq:2.16}) with Eq.\ (\ref{eq:2.8}) is given 
by \cite{Garretson92} 
\begin{eqnarray} 
\phi 
\Eqn{\approx}
\phi_1 \exp \left[ -\frac{3}{2} H_{\mathrm{inf}} \left( t-t_1 \right)  
\right] \sin \left[ m \left( t-t_1 \right) + \frac{\pi}{2} \right].  
\label{eq:3.8} 
\end{eqnarray}
In deriving Eq.\ (\ref{eq:3.8}), we have used Eq.\ (\ref{eq:3.1}).  
\if
Moreover, in deriving Eq.\ (\ref{eq:3.9}), we have solved 
Eq.\ (\ref{eq:2.16}) by neglecting $\ddot{\phi}$ because 
$\left| \ddot{\phi} / 
\left( 3 H_{\mathrm{inf}} \dot{\phi} \right) \right| \ll 1$ for 
$m \ll H_{\mathrm{inf}}$.  
In fact, using the solution in Eq.\ (\ref{eq:3.9}), we find that 
$\left| \ddot{\phi}/ \left( 3 H_{\mathrm{inf}} \dot{\phi} \right) 
\right| = \left[ m / \left( 3H_{\mathrm{inf}} \right) \right]^2 \ll 1$.  
\fi
Moreover, we consider the case in which after inflation, 
the pseudoscalar field $\phi$ decays in the radiation-dominated stage 
and hence the entropy per comoving volume remains practically constant.

\subsection{Evolution of the $U(1)_Y$ gauge field}

Next, we consider the evolution of the $U(1)_Y$ gauge field.  
To begin with, we shall quantize the $U(1)_Y$ gauge field 
$Y_{\mu}(t,\Vec{x})$.  
It follows from the hypercharge electromagnetic part of our model Lagrangian 
in Eq.\ (\ref{eq:2.5}) that the canonical momenta conjugate to 
$Y_{\mu}(t,\Vec{x})$ are given by 
\begin{eqnarray}
{\pi}_0 = 0, \hspace{5mm} {\pi}_{i} = f(\Phi) a(t) \dot{Y_i}(t,\Vec{x}).
\label{eq:3.9} 
\end{eqnarray}
We impose the canonical commutation relation 
between $Y_i(t,\Vec{x})$ and ${\pi}_{j}(t,\Vec{x})$, 
\begin{eqnarray} 
  \left[ \hspace{0.5mm} Y_i(t,\Vec{x}), {\pi}_{j}(t,\Vec{y}) 
  \hspace{0.5mm} \right] = i
 \int \frac{d^3 k}{{(2\pi)}^{3}}
             e^{i \Vecs{k} \cdot \left( \Vecs{x} - \Vecs{y} \right)}
        \left( {\delta}_{ij} - \frac{k_i k_j}{k^2 } \right),
\label{eq:3.10} 
\end{eqnarray}
where $\Vec{k}$ is comoving wave number, and $k$ denotes its amplitude 
$|\Vec{k}|$.  
From this relation, we obtain the expression for $Y_i(t,\Vec{x})$ as 
\begin{eqnarray} 
  Y_i(t,\Vec{x}) = \int \frac{d^3 k}{{(2\pi)}^{3/2}}
  \left[ \hspace{0.5mm} \hat{b}(\Vec{k}) 
        Y_i(t,\Vec{k})e^{i \Vecs{k} \cdot \Vecs{x} }
       + {\hat{b}}^{\dagger}(\Vec{k})
       {Y_i^*}(t,\Vec{k})e^{-i \Vecs{k} \cdot \Vecs{x}} \hspace{0.5mm} \right],
\label{eq:3.11} 
\end{eqnarray}
where $\hat{b}(\Vec{k})$ and ${\hat{b}}^{\dagger}(\Vec{k})$ 
are the annihilation and creation operators which satisfy 
\begin{eqnarray} 
\left[ \hspace{0.5mm} \hat{b}(\Vec{k}), {\hat{b}}^{\dagger}({\Vec{k}}^{\prime}) \hspace{0.5mm} \right] = 
{\delta}^3 (\Vec{k}-{\Vec{k}}^{\prime}), \hspace{5mm}
\left[ \hspace{0.5mm} \hat{b}(\Vec{k}), \hat{b}({\Vec{k}}^{\prime})
\hspace{0.5mm} \right] = 
\left[ \hspace{0.5mm} 
{\hat{b}}^{\dagger}(\Vec{k}), {\hat{b}}^{\dagger}({\Vec{k}}^{\prime})
\hspace{0.5mm} \right] = 0.
\label{eq:3.12} 
\end{eqnarray}
It follows from Eqs.\ (\ref{eq:3.10}) and (\ref{eq:3.11}) that 
the normalization condition for $Y_i(k,t)$ reads 
\begin{eqnarray} 
Y_i(k,t){\dot{Y}}_j^{*}(k,t) - {\dot{Y}}_j(k,t){Y_i^{*}}(k,t)
= \frac{i}{f a} \left( {\delta}_{ij} - \frac{k_i k_j}{k^2 } \right).
\label{eq:3.13} 
\end{eqnarray}

From now on we choose the $x^3$ axis to lie along the spatial momentum 
direction \Vec{k} and denote the transverse directions $x^{I}$ with 
$I=1, 2$.  From Eq.\ (\ref{eq:2.22}), 
we find that the Fourier modes $Y_I(k,t)$ of 
the $U(1)_Y$ gauge field  satisfy the following equations:  
\begin{eqnarray}
\ddot{Y}_1(k,t)
+ \left( H_{\mathrm{inf}} + \frac{\dot{f}}{f} 
\right) \dot{Y}_1(k,t)
+ \frac{k^2}{a^2} Y_1(k,t) 
+ik \frac{g_{\mathrm{ps}}}{M} 
\frac{1}{af} \frac{d \left(f \phi \right)}{d t} 
Y_2 (k,t) \Eqn{=} 0, 
\label{eq:3.14}  \\[3mm] 
\ddot{Y}_2(k,t) 
+ \left( H_{\mathrm{inf}} + \frac{\dot{f}}{f} 
\right) \dot{Y}_2(k,t) 
+ \frac{k^2}{a^2} Y_2(k,t) 
-ik \frac{g_{\mathrm{ps}}}{M} 
\frac{1}{af} \frac{d \left(f \phi \right)}{d t} 
Y_1(k,t)  \Eqn{=} 0.  
\label{eq:3.15}
\end{eqnarray}   
In order to decouple the system of Eqs.\ (\ref{eq:3.14}) and (\ref{eq:3.15}), 
we consider circular polarizations expressed by 
the combination of linear polarizations as 
$Y_{\pm}(k,t)  \equiv Y_1(k,t) \pm i Y_2(k,t)$.  
From Eqs.\ (\ref{eq:3.14}) and (\ref{eq:3.15}), we find that 
$Y_{\pm}(k,t)$ satisfies the following equation:  
\begin{eqnarray}
\ddot{Y}_{\pm}(k,t) 
+ \left( H_{\mathrm{inf}} + \frac{\dot{f}}{f} 
\right) \dot{Y}_{\pm}(k,t) 
+ \left[ 
\left( \frac{k}{a} \right)^2 
\pm \frac{g_{\mathrm{ps}}}{M} 
\left( \frac{\dot{f}}{f} \phi + \dot{\phi} \right) 
\left( \frac{k}{a} \right)
\right] Y_{\pm}(k,t) = 0.  
\label{eq:3.16}
\end{eqnarray}   

Since it is difficult to obtain the analytic solution of Eq.\ (\ref{eq:3.16}), 
we numerically solve this equation during inflation.  
Here we assume that the initial amplitude of $Y_+(k,t)$ and that of $Y_-(k,t)$ 
are the same value and we take the time $t_1$ as the initial time.  
In the inflationary stage, 
the difference between the evolution of $Y_+(k,t)$ and that of $Y_-(k,t)$ 
is induced by the coupling between the pseudoscalar field and 
hypercharge electromagnetic fields and thus the hypermagnetic helicity is 
generated \cite{Field00, BO99, Giovannini00, Campanelli05}.   
During inflation ($t_1 \leq t \leq t_{\mathrm{R}}$), 
the amplitude of $Y_{\pm}(k,t)$ is expressed as 
\begin{eqnarray} 
Y_{\pm}(k,t) = C_{\pm}(t) Y_{\pm}(k,t_1), 
\label{eq:3.17}
\end{eqnarray}   
where $C_{\pm}(t)$ is a numerical value 
obtained by numerical calculations and we take $C_{\pm}(t_1)=1$.  

In order to obtain the initial amplitude of $Y_\pm(k,t)$, 
we here consider the solution of Eq.\ (\ref{eq:3.16}) inside of the 
horizon, i.e., on subhorizon scales, $k/(aH) \gg 1$.  
Replacing the independent variable $t$ by $\eta$, we find that 
in the short-wavelength 
limit, $k \rightarrow \infty$, 
Eq.\ (\ref{eq:3.16}) is approximately given by 
\begin{eqnarray}
Y_{\pm}^{\prime \prime}(k,\eta) + 
\frac{f^{\prime}}{f} Y_{\pm}^{\prime}(k,\eta) 
+ k^2 Y_{\pm}(k,\eta) = 0,  
\label{eq:3.18}
\end{eqnarray}
where 
the prime denotes differentiation with respect to the conformal time $\eta$.  
Here, in deriving Eq.\ (\ref{eq:3.18}), we have taken only the term 
proportional to $k^2$ on the right-hand side of Eq.\ (\ref{eq:3.16}) and 
neglected that proportional to $k$.  

The inside solution is given by 
\begin{eqnarray}
Y_{\pm}^{\mathrm{in}} (k,\eta) = 
\frac{1}{\sqrt{2k}} f^{-1/2} e^{-ik\eta}, 
\label{eq:3.19} 
\end{eqnarray} 
where we have determined the coefficient of this solution by requiring that 
the vacuum reduces to the one in Minkowski spacetime in the short-wavelength 
limit.  
In fact, using Eq.\ (\ref{eq:3.19}), we find 
\begin{eqnarray}
{Y_{\pm}^{\mathrm{in}}}^{\prime \prime}(k,\eta) + 
\frac{f^{\prime}}{f} {Y_{\pm}^{\mathrm{in}}}^{\prime}(k,\eta) 
\Eqn{=} 
\left\{ - k^2 
- \frac{1}{2} \left[ -\frac{1}{2} \left( \frac{f^{\prime}}{f} \right)^2 + 
\frac{f^{\prime \prime}}{f}
\right] \right\} Y_{\pm}^{\mathrm{in}}(k,\eta)  
\label{eq:3.20} \\[3mm] 
\Eqn{\approx} 
- k^2 Y_{\pm}^{\mathrm{in}} (k,\eta),  
\label{eq:3.21}
\end{eqnarray}
where the approximate equality in Eq.\ (\ref{eq:3.21}) follows from 
$-k \eta \gg 1$.  Hence we see that 
the inside solution in Eq.\ (\ref{eq:3.19}) approximately satisfies 
Eq.\ (\ref{eq:3.18}).  

Here we assume that the initial amplitude of $Y_{\pm}(k,t)$ 
is approximately given by the inside solution in Eq.\ (\ref{eq:3.19}).  
Hence the initial amplitude of $Y_{\pm}(k,t)$ at the time $t_1$ is given by 
\begin{eqnarray}
|Y_{\pm}(k,t_1)| \approx \frac{1}{\sqrt{2k f(t_1)}}. 
\label{eq:3.22} 
\end{eqnarray} 

The numerical results of the evolution of $C_{\pm}(t)$ are shown in
Figs.~1 and 2.  
Fig.~1 depicts the case in which 
$H_{\mathrm{inf}} = 10^{10}$GeV, $m = 10^9$GeV, $\lambda = - 133$, 
$\tilde{\lambda}=1.0$, 
$w=1/(75)$, $\phi_1 = 10^9$GeV, $M = 10^9$GeV, and 
$g_{\mathrm{ps}} = 1.0$ 
(the case (ii) in Table \ref{table:1} shown in the next subsection).  
On the other hand, 
Fig.~2 depicts the case in which 
$H_{\mathrm{inf}} = 10^{10}$GeV, $m = 10^{12}$GeV, 
$\lambda = - 135$, 
$\tilde{\lambda}=1.0$
$w=1/(75)$, $\phi_1 = 10^{12}$GeV, $M = 10^{12}$GeV, and 
$g_{\mathrm{ps}} = 1.0$ (the case (iii) in Table \ref{table:1}).  
In Figs.~1 and 2, 
we have used the evolution of $\phi$ in the positive sign equation in 
(\ref{eq:3.7}) and that in Eq.\ (\ref{eq:3.8}), respectively.  
Moreover, we have used 
$k/a = \exp \left[ - H_{\mathrm{inf}} \left( t-t_1\right)\right] 
H_{\mathrm{inf}}$, which follows from Eq.\ (\ref{eq:3.1}) and 
$k/\left( a_1 H_{\mathrm{inf}}\right)=1$.  
The solid  curves represents $C_+(t)$ and the dotted curves 
represents $C_-(t)$.  
Here we have started to calculate Eq.\ (\ref{eq:3.16}) numerically at 
$t=t_1= H_{\mathrm{inf}}^{-1}$ and we assume that the initial value 
of $C_\pm(t)$ at $t=t_1$ is $C_\pm(t_1) = 1.0$.   
From Figs.\ 1 and 2, we understand that 
after several Hubble expansion times, 
both the value of $C_+(t)$ and that of $C_-(t)$ becomes almost constant.  
This qualitative behavior is common to the case $ m < H_{\mathrm{inf}}$ 
and the case $m \gg H_{\mathrm{inf}}$.  
Finally, we note that if we take a larger value of $g_{\mathrm{ps}}$ 
and/or that of $\phi_1/M$ (in all the cases in Table \ref{table:1} 
we have taken $\phi_1/M = 1.0$), the difference between 
the evolution and amplitude of $C_+(t)$ and those of $C_-(t)$ becomes larger.

\subsection{Chern-Simons number density}

In this subsection, we consider the generation of the 
Chern-Simons number density in the inflationary stage and 
estimate the resultant baryon asymmetry.  
The proper hyperelectric and hypermagnetic fields are given by \cite{Ratra}
\begin{eqnarray}
\Eqn{} {E_Y}_i^{\mathrm{proper}}(t,\Vec{x})
    = a^{-1}{E_Y}_i(t,\Vec{x}) = -a^{-1}\dot{Y}_i(t,\Vec{x}), 
\label{eq:3.23} \\[3mm]
\Eqn{} {B_Y}_i^{\mathrm{proper}}(t,\Vec{x})
    = a^{-1}{B_Y}_i(t,\Vec{x}) = 
a^{-2}{\epsilon}_{ijk}{\partial}_j Y_k(t,\Vec{x}),
\label{eq:3.24}    
\end{eqnarray} 
where ${E_Y}_i(t,\Vec{x})$ and ${B_Y}_i(t,\Vec{x})$ are 
the comoving hyperelectric and hypermagnetic fields, 
and ${\epsilon}_{ijk}$ is the totally antisymmetric tensor
(\hspace{0.5mm}${\epsilon}_{123}=1$\hspace{0.5mm}).  

The density of the baryonic number $n_\mathrm{B}$ is given by \cite{GS98-2} 
\begin{eqnarray} 
n_\mathrm{B} \Eqn{=} -\frac{n_\mathrm{f}}{2} \Delta n_\mathrm{CS}, 
\label{eq:3.25} \\[3mm]   
\Delta n_\mathrm{CS} \Eqn{=} - 
\frac{g^{\prime \hspace{0.3mm} 2}}{4 \pi^2} 
\int^{t} {\Vec{E}}_Y \cdot {\Vec{B}}_Y d \tilde{t}.  
\label{eq:3.26}
\end{eqnarray} 
Here, $n_\mathrm{f}$ is the number of fermionic generations 
(throughout this paper we use $n_\mathrm{f} = 3$) 
and 
$\Delta n_\mathrm{CS}$ is the Chern-Simons number density.  

It follows from Eqs.\ (\ref{eq:3.23}) and (\ref{eq:3.24}) that 
the Fourier modes ${E_Y}_\pm^{\mathrm{proper}}(k,t) = 
{E_Y}_1^{\mathrm{proper}}(k,t) \pm i {E_Y}_2^{\mathrm{proper}}(k,t)$ 
and ${B_Y}_\pm^{\mathrm{proper}}(k,t) = 
{B_Y}_1^{\mathrm{proper}}(k,t) \pm i {B_Y}_2^{\mathrm{proper}}(k,t)$ 
satisfy the following relations:  
\begin{eqnarray}  
{E_Y}_\pm^{\mathrm{proper}}(k,t) \Eqn{=}  
\pm \frac{1}{k}  \frac{\partial {B_Y}_\pm^{\mathrm{proper}}(k,t)}{\partial t}, 
\label{eq:3.27} \\[3mm] 
{E_Y}_\pm^{\mathrm{proper}}(k,t) {B_Y}_\pm^{\mathrm{proper}}(k,t) \Eqn{=}  
\pm \frac{1}{2}\frac{1}{k} \frac{\partial 
\left[ {{B_Y}_\pm^{\mathrm{proper}}}(k,t) \right]^2}{\partial t}.  
\label{eq:3.28} 
\end{eqnarray} 

On the other hand, 
the energy density of the proper hypermagnetic field 
in Fourier space is given by  
\begin{eqnarray}
{\rho}_{B_Y}(k,t) 
= 
\frac{1}{2}
\left[  
\left| {B_Y}_+^{\mathrm{proper}}(k,t)
\right|^2 
+
\left| {B_Y}_-^{\mathrm{proper}}(k,t)
\right|^2 \right] f, 
\label{eq:3.29}  
\end{eqnarray} 
\begin{eqnarray}
\left| {B_Y}_\pm^{\mathrm{proper}}(k,t)
\right|^2 
= 
\frac{1}{a^2} \left( \frac{k}{a} \right)^2 
 |Y_\pm(k,t)|^2.  
\label{eq:3.30} 
\end{eqnarray} 
In deriving Eq.\ (\ref{eq:3.30}), we have used Eq.\ (\ref{eq:3.24}).  
Multiplying ${\rho}_{B_Y}(k,t)$ by phase-space density:\ $4\pi k^3/(2\pi)^3$, 
we obtain the energy density of the proper magnetic field 
in the position space 
\begin{eqnarray}
{\rho}_{B_Y}(L,t) = 
 \frac{k^3}{4{\pi}^2}
\left[  
\left| {B_Y}_+^{\mathrm{proper}}(k,t)
\right|^2 
+
\left| {B_Y}_-^{\mathrm{proper}}(k,t)
\right|^2 \right] f, 
\label{eq:3.31}
\end{eqnarray} 
on a comoving scale $L=2\pi/k$.

Using Eqs.\ (\ref{eq:3.17}), (\ref{eq:3.26}), (\ref{eq:3.28}), 
(\ref{eq:3.30}), and (\ref{eq:3.31}), we find that 
the Chern-Simons number density in the inflationary stage is given by 
\begin{eqnarray}   
\Delta n_\mathrm{CS} \Eqn{=} 
- \frac{g^{\prime \hspace{0.3mm} 2}}{4 \pi^2} 
\frac{1}{k} \frac{1}{f} {\rho}_{B_Y}(L,t) \mathcal{A}(t),  
\label{eq:3.32} \\[3mm] 
\mathcal{A}(t) \Eqn{=} 
\frac{|C_+(t)|^2 - |C_-(t)|^2}{|C_+(t)|^2 + |C_-(t)|^2}. 
\label{eq:3.33} 
\end{eqnarray}

Here we consider the case in which 
after inflation the Universe is reheated immediately at $t=t_\mathrm{R}$.  
Moreover, we assume that the instantaneous reheating stage 
is much before the EWPT 
(the background temperature at the EWPT is $T_\mathrm{EW} \sim 100$GeV).  
The conductivity of the Universe ${\sigma}_\mathrm{c}$ 
is negligibly small during inflation, because there are few charged particles 
at that time. 
After reheating, however, a number of charged particles are produced, 
so that the conductivity immediately jumps to a large value:\ 
${\sigma}_\mathrm{c} \gg H \hspace{1.5mm} 
(\hspace{0.5mm}t \geq t_\mathrm{R}\hspace{0.5mm})$.  
Consequently, 
for a large enough conductivity at the instantaneous reheating stage, 
hyperelectric fields accelerate charged particles and dissipate.  
On the other hand, the proper hypermagnetic fields evolve in 
proportion to $a^{-2}(t)$ in the radiation-dominated stage 
and the subsequent matter-dominated stage 
(\hspace{0.5mm}$t \geq t_\mathrm{R}$\hspace{0.5mm}) \cite{Ratra, Bamba1}.  
Furthermore, the hypermagnetic helicity, i.e., the Chern-Simons number, is 
conserved \cite{GS98-2, EParker, Biskamp}.  
The Chern-Simons number will be released at the EWPT in the form of fermions, 
which will not be destroyed by the 
sphaleron processes \cite{Sphaleron} if the EWPT is 
strongly first order \cite{GS98-1, GS98-2}.  Although the EWPT is 
not strongly first order in the MSM, 
the EWPT could be strongly first order 
in the extensions of the MSM \cite{Funakubo05}.   
Moreover, at the EWPT the hypermagnetic fields are replaced by the ordinary 
magnetic fields \cite{GS98-1, GS98-2}.  

Moreover, we discuss the finite conductivity effects.  
We consider the case in which in the last stage of inflation 
the conductivity has a finite value.  In this case,  
the term of the conductivity ${\sigma}_\mathrm{c}$ is added to the inside of 
the parentheses in the second term on the right-hand side of 
Eq.\ (\ref{eq:3.16}), which is the coefficient of 
$\dot{Y}_{\pm}(k,t)$.  On the other hand, 
there also exists the term of $\dot{f}/f$ in the inside of 
the parentheses in the second term on the right-hand side of 
Eq.\ (\ref{eq:3.16}).  Hence the term of $\dot{f}/f$ has the same 
effect as the finite conductivity ${\sigma}_\mathrm{c}$.  
As the physical effect, the term of $\dot{f}/f$ makes 
both the value of $C_+(t)$ and that of $C_-(t)$ in Eq.\ (\ref{eq:3.17}) 
becomes almost constant after several Hubble expansion times.  
In fact, in the case in which in the last stage of inflation 
the conductivity ${\sigma}_\mathrm{c}$ has a finite value, e.g., 
${\sigma}_\mathrm{c} = 100 H_{\mathrm{inf}}$, we have numerically calculated 
the evolution of $C_+(t)$ and $C_-(t)$.  As a result, we have found that 
the results in the case in which we take into account the finite conductivity 
effects are almost same as the results without taking into account the finite 
conductivity effects.  This is because 
there exists the term of $\dot{f}/f$ 
in the coefficient of $\dot{Y}_{\pm}(k,t)$ in Eq.\ (\ref{eq:3.16}) and 
this term has the same physical effect as the finite 
conductivity ${\sigma}_\mathrm{c}$.  
In the models in Refs.~\cite{BO99, Giovannini00}, the finite conductivity 
effects have a influence on results because in these models the coupling 
$f(\Phi)$ between the dilaton and the hypercharge electromagnetic fields 
is not considered and hence there does not exist the term of $\dot{f}/f$ 
in the coefficient of $\dot{Y}_{\pm}(k,t)$.  In our model, however, 
since we consider the coupling $f(\Phi)$, 
there exists the term of $\dot{f}/f$ in the coefficient of 
$\dot{Y}_{\pm}(k,t)$ in Eq.\ (\ref{eq:3.16}), which 
has the same physical effect as the finite 
conductivity ${\sigma}_\mathrm{c}$.  
Consequently, in our model, the finite conductivity 
effects have little influence on results.  
Thus we think that the assumption that 
after reheating the conductivity immediately jumps to a large value, 
${\sigma}_\mathrm{c}\gg H \hspace{1.5mm} 
(\hspace{0.5mm}t \geq t_\mathrm{R}\hspace{0.5mm})$, is appropriate, 
and that the results under this assumption are proper.  

It follows from Eqs.\ (\ref{eq:3.25}), (\ref{eq:3.32}), 
and (\ref{eq:3.33}) that after the EWPT, 
the ratio of the density of the baryonic number $n_\mathrm{B}$ 
to the entropy density $s$ is given by 
\begin{eqnarray}    
\frac{n_\mathrm{B}}{s} \Eqn{=} 
n_\mathrm{f} \frac{g^{\prime \hspace{0.3mm} 2}}{8\pi^2} 
\frac{1}{k} \frac{\rho_B (L,t)}{s} \mathcal{A}(t_\mathrm{R}),  
\label{eq:3.34} 
\end{eqnarray} 
with 
\begin{eqnarray}    
\rho_B (L,t) \Eqn{=} 
\frac{1}{8\pi^2} \frac{1}{f(t_1)} \left( \frac{k}{a} \right)^4 
\left[ |C_+(t_\mathrm{R})|^2 + |C_-(t_\mathrm{R})|^2 \right],
\label{eq:3.35}  \\[3mm]
f(t_1) \Eqn{=} \left( 1 - \tilde{\lambda}^2 w N \right)^{-X}, 
\label{eq:3.36}  \\[3mm]
N \Eqn{=} H_{\mathrm{inf}} \left( t_\mathrm{R} - t_1 \right), 
\label{eq:3.37}  \\[3mm]
X \Eqn{\equiv} \frac{\lambda}{\tilde{\lambda}}, 
\label{eq:3.38}
\end{eqnarray}   
where 
we have used the fact that when $t \geq t_\mathrm{R}$, 
$f = 1$, and that after the EWPT, 
$\rho_{B_Y} (L,t) \rightarrow \rho_B (L,t)$, where 
$\rho_B (L,t)$ is the energy density of the ordinary magnetic fields.  
Moreover, we have taken into account the fact that 
for a large enough conductivity after the instantaneous reheating stage, 
the proper hypermagnetic fields evolve in proportion to $a^{-2}(t)$ 
as stated above.  
Here, $N$ is the number of $e$-folds between the time $t_1$
and the end of inflation $t_{\mathrm{R}}$.  
Furthermore, in deriving Eq.\ (\ref{eq:3.36}), we have used 
Eqs.\ (\ref{eq:2.6}) and (\ref{eq:3.4}) with $\Phi_{\mathrm{R}} = 0$.  

In order to estimate the value of $n_\mathrm{B}/s$, 
we use the following relations \cite{Kolb}:   
\begin{eqnarray}
H_{0} \Eqn{=} 100 h 
\hspace{1mm} \mathrm{km} \hspace{1mm} {\mathrm{s}}^{-1} \hspace{1mm} 
{\mathrm{Mpc}}^{-1} 
= 2.1 h \times 10^{-42} {\mathrm{GeV}},  
\label{eq:3.39} \\[3mm] 
{\rho}_{\varphi} \left( t_{\mathrm{R}} \right) \Eqn{=} 
\frac{{\pi}^2}{30} g_\mathrm{R} {T_\mathrm{R}}^4 \hspace{3mm} 
\left( g_{\mathrm{R}} \approx 200 \right),  
\label{eq:3.40} \\[3mm] 
N \Eqn{=} 45 + \ln \left( \frac{L}{\mathrm{[Mpc]}} \right) + 
    \ln \left\{ \frac{ \left[ 30/({\pi}^2 g_\mathrm{R} )  \right]^{1/12} 
               {{\rho}_{\varphi}}^{1/4} }
          {10^{38/3} \hspace{1mm} \mathrm{[GeV]}}  \right\},
\label{eq:3.41} \\[3mm] 
\frac{a_{\mathrm{R}}}{a_0}  
\Eqn{=} 
\left( \frac{ g_{\mathrm{R}} }{3.91} \right)^{-1/3}
\frac{T_{ \gamma 0} }{ T_{\mathrm{R}} }  
\hspace{0mm} \approx \hspace{0mm}
\frac{2.35\times10^{-13} \hspace{1mm} [\mathrm{GeV}]}{ 3.71 T_{\mathrm{R}}}
   \hspace{3mm} \left( 
T_{ \gamma 0} \approx 2.73 \hspace{1mm} \mathrm{K} \right), 
\label{eq:3.42}  \\[3mm] 
s_0 \Eqn{=} 2.97 \times 10^{3} 
\left( \frac{T_{ \gamma 0}}{2.75 \hspace{1mm} [\mathrm{K}]} \right)^3 
{\mathrm{cm}}^{-3},
\label{eq:3.43}
\end{eqnarray} 
where $H_{0}$
is the Hubble constant at 
the present time (throughout this paper we use $h=0.70$ \cite{HST}), 
$g_{\mathrm{R}}$ is the total number of degrees of freedom for 
relativistic particles at the reheating epoch, 
$T_{\mathrm{R}}$ is reheating temperature,  
$T_{\gamma 0}$ is the present temperature of 
the cosmic microwave background (CMB) radiation, 
and $s_0$ is the entropy density at the present time. 
Moreover, we use $g^{\prime \hspace{0.3mm} 2}/(4\pi) = 
\alpha_{\mathrm{EM}}/ \cos^2 \theta_{\mathrm{w}}$, where 
$\alpha_{\mathrm{EM}} = 1/(137)$ is the fine-structure constant and 
$\theta_{\mathrm{w}}$ is the weak mixing angle.  The experimentally measured 
value of $\theta_{\mathrm{w}}$ 
is given by $\sin^2 \theta_{\mathrm{w}} \simeq 0.23$ 
\cite{Trodden99}.


\begin{table}[tbp]
\caption{
Estimates of the value of $n_\mathrm{B}/s$ 
for $w=1/(75)$, $\phi_1 = M =m$, $g_{\mathrm{ps}} = 1.0$, and 
$\tilde{\lambda}=1.0$.  
Here we have used the evolution of $\phi$ 
in the positive sign equation in (\ref{eq:3.7}) 
for the cases (i), (ii), and (iv), 
and that in Eq.\ (\ref{eq:3.8}) for those (iii) and (v).  
}
\begin{center}
\tabcolsep = 2mm 
\begin{tabular}
{cccccccc}
\hline
\hline
& $\left| n_\mathrm{B}/s \right |$  
& $B(H_0^{-1},t_0) \hspace{1mm} [\mathrm{G}]$  
& $H_{\mathrm{inf}} \hspace{1mm} [\mathrm{GeV}]$   
& $m \hspace{1mm} [\mathrm{GeV}]$  
& $C_+(t_{\mathrm{R}})$ 
& $C_-(t_{\mathrm{R}})$ 
& $X$
\\[0mm]
\hline
(i) 
& $3.6 \times 10^{-10}$
& $2.7 \times 10^{-24}$
& $1.0 \times 10^{14}$
& $1.0 \times 10^{12}$
& $0.367$
& $2.23$
& $-1.11 \times 10^2$
\\[0mm]
(ii) 
& $1.0 \times 10^{-10}$
& $1.4 \times 10^{-24}$
& $1.0 \times 10^{10}$
& $1.0 \times 10^{9}$
& $0.369$
& $2.18$
& $-1.33 \times 10^2$
\\[0mm]
(iii) 
& $1.5 \times 10^{-10}$
& $3.5 \times 10^{-24}$
& $1.0 \times 10^{10}$
& $1.0 \times 10^{12}$
& $1.04$
& $0.816$
& $-1.35 \times 10^2$
\\[0mm]
(iv) 
& $0.96 \times 10^{-10}$
& $1.4 \times 10^{-24}$
& $1.0 \times 10^{5}$
& $1.0 \times 10^{3}$
& $0.368$
& $2.17$
& $-1.65 \times 10^2$
\\[0mm]
(v) 
& $2.8 \times 10^{-10}$
& $4.7 \times 10^{-24}$
& $1.0 \times 10^{5}$
& $1.0 \times 10^{7}$
& $1.04$
& $0.808$
& $-1.68 \times 10^2$
\\[1mm]
\hline
\hline
\end{tabular}
\end{center}
\label{table:1}
\end{table}


Consequently, from Eq.\ (\ref{eq:3.34}), we can estimate the 
value of the ratio of the density of the baryonic number $n_\mathrm{B}$ 
to the entropy density $s$, which is observationally estimated as 
$n_\mathrm{B}/s = 0.92 \times 10^{-10}$ 
by using the 
the first year Wilkinson Microwave Anisotropy Probe 
(WMAP) data on the anisotropy of the CMB radiation \cite{Spergel}.  
Table \ref{table:1} displays the estimate of the value of $n_\mathrm{B}/s$ 
for $w=1/(75)$, $\phi_1 = M =m$, $g_{\mathrm{ps}} = 1.0$, and 
$\tilde{\lambda}=1.0$.  
Here we have used the evolution of $\phi$ 
in the positive sign equation in (\ref{eq:3.7}) 
for the cases (i), (ii), and (iv), 
and that in Eq.\ (\ref{eq:3.8}) for those (iii) and (v).  
In Table \ref{table:1}, we have considered the BAU 
induced by the helicity of 
the hypermagnetic fields whose present scale is the present horizon scale, 
$L = H_0^{-1}$.  Hence the resultant BAU is 
homogeneous over the present horizon.   
Moreover, 
in estimating the value of the energy density of the magnetic fields 
in Eq.\ (\ref{eq:3.35}), we have used $a_{\mathrm{R}}/a_1 = \exp (N)$ and 
$k / \left( a_1 H_{\mathrm{inf}} \right) = 1$.  
Furthermore, 
a constraint on $H_\mathrm{inf}$ 
from tensor perturbations \cite{Abbott1, Rubakov} is obtained by 
using the WMAP three year data on temperature fluctuations \cite{Spergel06},
$H_\mathrm{inf} < 5.9 \times 10^{14}$GeV.  

From Table \ref{table:1}, we see that if the magnetic fields with 
the field strength $\sim 10^{-24}$G on the horizon scale at the 
present time are generated, the resultant value of $n_\mathrm{B}/s$ 
can be as large as $10^{-10}$, which is consistent with the above 
observational estimation of WMAP.  
When the generated magnetic fields with the field 
strength $\sim 10^{-24}$G on the horizon scale at the present time 
play the role of 
seed magnetic fields of galactic magnetic fields, 
by assuming the magnetic flux conservation, $B r^2 = \mathrm{constant}$, 
where $r$ is a scale, and using the scale ratio 
$r_\mathrm{gal}/H_0^{-1} \sim 10^{-5}$, 
where $r_\mathrm{gal}$ is the scale of galaxies, 
we can estimate the seed magnetic field strength of 
the galactic magnetic fields 
as $B_\mathrm{gal}^{(\mathrm{seed})} \sim 10^{-14}~\mathrm{G}$ which 
is sufficient for the following dynamo amplification.  
Thus, in this way, the main problem of cosmological magnetic fields 
can be solved by obtaining both large spatial scale and 
large magnetic field amplitude 
as it should be for the relic seed magnetic fields obeying dynamo theories.  

Here we state the reason why in this model 
the amplitude of the generated magnetic fields can be as large as 
$\sim 10^{-24}$G on the horizon scale at the present time.  
The reason is that the conformal invariance of the 
hypercharge electromagnetic fields is extremely broken 
through the coupling between the dilaton and 
the hypercharge electromagnetic fields by introducing 
a huge hierarchy 
between the coupling constant of the dilaton to 
the hypercharge electromagnetic fields 
$\lambda$ and the coupling constant $\tilde\lambda$ of the dilaton potential, 
$|X| = \left|\lambda/\tilde{\lambda}\right| \gg 1$. 
From Table \ref{table:1}, we see that 
in order that 
the magnetic fields with 
the field strength $\sim 10^{-24}$G on the horizon scale at the 
present time could be generated, 
we have to introduce the huge hierarchy $|X| \gg 1$.  
It follows from Eqs.\ (\ref{eq:3.22}) and 
(\ref{eq:3.36}) that if $|X| \gg 1$, the value of $f(t_1)$ is very small 
and hence the amplitude of 
$|Y_{\pm}(k,t_1)|$ is sufficiently large.  
Moreover, it follows from Figs. 1 and 2 that 
after several Hubble expansion times, 
both the value of $C_+(t)$ and that of $C_-(t)$ becomes almost constant.  
Hence it follows from Eq.\ (\ref{eq:3.17}) that 
the amplitude of $|Y_{\pm}(k,t)|$ is sufficiently large.  
Consequently, it follows from Eqs.\ (\ref{eq:3.30}), (\ref{eq:3.31}), and
(\ref{eq:3.35}) that the energy density of the generated magnetic fields, 
i.e., the resultant magnetic field amplitude can be as large as 
$\sim 10^{-24}$G on the horizon scale at the present time.  
This is the reason why the present amplitude of the generated magnetic fields 
in this model is larger than that in other inflation scenarios 
\cite{Scalar}.  

The physical reason why the coupling of the dilaton to the 
hypercharge electromagnetic fields has to be much stronger than 
the coupling in the dilaton potential in order that the amplitude of the 
generated magnetic fields could be as large as $\sim 10^{-24}$G on the horizon 
scale at the present time is as follows: 
In order to generate the magnetic fields with the sufficient strength, 
the conformal invariance of the hypercharge electromagnetic fields has to be 
extremely broken through the coupling $f(\Phi)$ between the dilaton and 
the hypercharge electromagnetic fields.  In order that this condition could 
be met, it is necessary that the change of 
the value of $f(\Phi)$ in the inflationary stage is much larger than 
the change of the value of the dilaton potential $V[\Phi]$ at that stage, 
in other words, $f(\Phi)$ evolves more rapidly than the dilaton 
potential $V[\Phi]$.  
Thus the coupling of the dilaton to the hypercharge electromagnetic 
fields has to be much stronger than the coupling in the dilaton potential, 
i.e., the value of $|\lambda|$ has to be much larger than that of 
$\tilde\lambda$.  
In this case, the value of $f(\Phi)$ can change from $f(t_1) \ll 1$ to 
$f(t_\mathrm{R}) = 1$ in the inflationary stage 
($t_1 \leq t \leq t_{\mathrm{R}}$).  
Consequently, from Eqs.\ (\ref{eq:3.35}) and (\ref{eq:3.36}), we see that 
the amplitude of the generated magnetic fields can be sufficiently large 
because $f(t_1)$ is much smaller than unity.  

\if
Finally, we note that in this scenario 
there remain some difficulties, e.g., 
the introduction of the unknown pseudoscalar field with a heavy 
mass and the necessary condition that the EWPT is strongly first order 
while the corresponding light Higgs mass is still an open problem.  
These difficulties should be solved in future work.  
\fi

\if
Finally, we note the following point:  
From Table \ref{table:1}, we see that 
in order that 
$n_\mathrm{B}/s$ 
could be as large as $10^{-10}$, in other words, 
the magnetic fields with 
the field strength $\sim 10^{-24}$G on the horizon scale at the 
present time could be generated, 
we have to introduce a huge hierarchy 
between the coupling constant of the dilaton to 
the hypercharge electromagnetic fields 
$\lambda$ and the coupling one $\tilde\lambda$ of the dilaton potential, 
$|X| = \left|\lambda/\tilde{\lambda}\right| \gg 1$.  
The reason is that the conformal invariance of the 
hypercharge electromagnetic fields 
must be broken extremely through the coupling between the dilaton and 
the hypercharge electromagnetic fields in order that the magnetic fields 
with sufficient strength to account for the observed value of 
$n_\mathrm{B}/s$ could be generated.  
The existence of the above huge hierarchy, however, seems to be unnatural 
in realistic high energy theories.  
The existence of the above huge hierarchy $|X| \gg 1$ seems to be unnatural 
in realistic high energy theories.  
In Ref.~\cite{Bamba2}, therefore, 
the present author and Yokoyama have discussed 
a possible solution to the above huge 
hierarchy between $\lambda$ and $\tilde\lambda$,  
by taking into account the effects of 
the stringy spacetime uncertainty relation (SSUR) \cite{Yoneya}.  
As a result, they have found that in power-law inflation models, 
owing to the consequences of the SSUR on metric perturbations, 
the magnetic fields with sufficient strength at the present time 
could be generated even in the case in which 
$\lambda$ and $\tilde\lambda$ are of the same order of magnitude.   
Furthermore, recently the present author and Sasaki have shown that 
if the conformal invariance of the Maxwell theory is broken through 
both the coupling of the dilaton to electromagnetic fields and 
that of the scalar curvature to those fields, 
large-scale magnetic fields with sufficient strength 
at the present time could be generated in models of power-law inflation 
without introducing the above huge hierarchy \cite{Bamba3}.  
\fi

\if
For example, we find that when the hypermagnetic fields on the 
horizon scale at the present time is generated, $n_\mathrm{B}/s$ at 
the present time is given by 
$n_\mathrm{B}/s \left(L = H_0^{-1} , t_0 \right)$ 
for $H_{\mathrm{inf}} = 10^{10}$GeV, $m = 10^{12}$GeV, 
$g_{\mathrm{ps}} = 1.0$, 
$\tilde{\lambda}=1.0$, $w=1/75$, $\phi_1 = M =m$, 
and $M = 10^{10}$GeV.  
\fi

\section{Conclusion}

In the present paper we have studied 
the generation of the BAU from the helicity of hypermagnetic fields 
in inflationary cosmology, 
taking into account the breaking of the conformal invariance of 
the hypercharge electromagnetic fields by introducing 
both a coupling with the dilaton and that with a pseudoscalar field.  
Owing to the coupling between the pseudoscalar field and 
the hypercharge electromagnetic fields, the hypermagnetic helicity, 
which corresponds to the Chern-Simons number, is 
generated in the inflationary stage.  
The Chern-Simons number 
stored in the hypercharge electromagnetic fields 
is converted into fermions at the EWPT due to the Abelian anomaly, 
and at the same time 
the hypermagnetic fields are replaced by the ordinary magnetic fields, 
which survive after the EWPT.  
The generated fermions can survive after the EWPT 
if the EWPT is strongly first order.  
In the extensions of the MSM, the above condition could be satisfied.  

As a result, we have found that 
if the magnetic fields with 
sufficient strength on the horizon scale at the 
present time are generated, the resultant value of the ratio of 
the density of the baryonic number to the entropy density, 
$n_\mathrm{B}/s$, 
can be as large as $10^{-10}$, which is consistent with the magnitude 
of the BAU suggested by observations obtained from WMAP.

\section*{Acknowledgements}
The author is deeply grateful to Mikhail Shaposhnikov, 
Jun'ichi Yokoyama, and Yasunari Kurita for helpful discussions.  
The author's work was partially supported by 
the Monbu-Kagaku Sho 21st century COE Program 
``Center for Diversity and Universality in Physics" 
and was also supported by 
a Grant-in-Aid provided by the 
Japan Society for the Promotion of Science.


\newpage


\begin{figure}[tbp]
\begin{center}
   \includegraphics{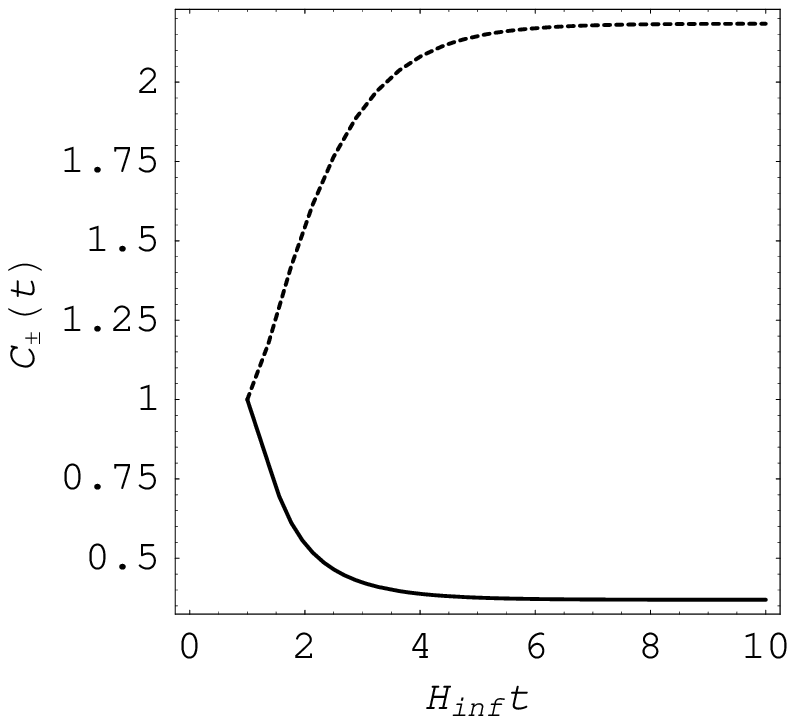}
\caption{
Numerical results of the evolution of $C_{\pm}(t)$ 
for $H_{\mathrm{inf}} = 10^{10}$GeV, $m = 10^9$GeV, $\lambda = - 133$, 
$\tilde{\lambda}=1.0$, 
$w=1/(75)$, $\phi_1 = 10^9$GeV, $M = 10^9$GeV, and 
$g_{\mathrm{ps}} = 1.0$ (the case (ii) in Table \ref{table:1}).  
Here 
we have used the evolution of $\phi$ in the positive sign equation in 
(\ref{eq:3.7}).  
The solid curves represents $C_+(t)$ and the dotted curves 
represents $C_-(t)$.   
}

\vspace{15mm}

   \includegraphics{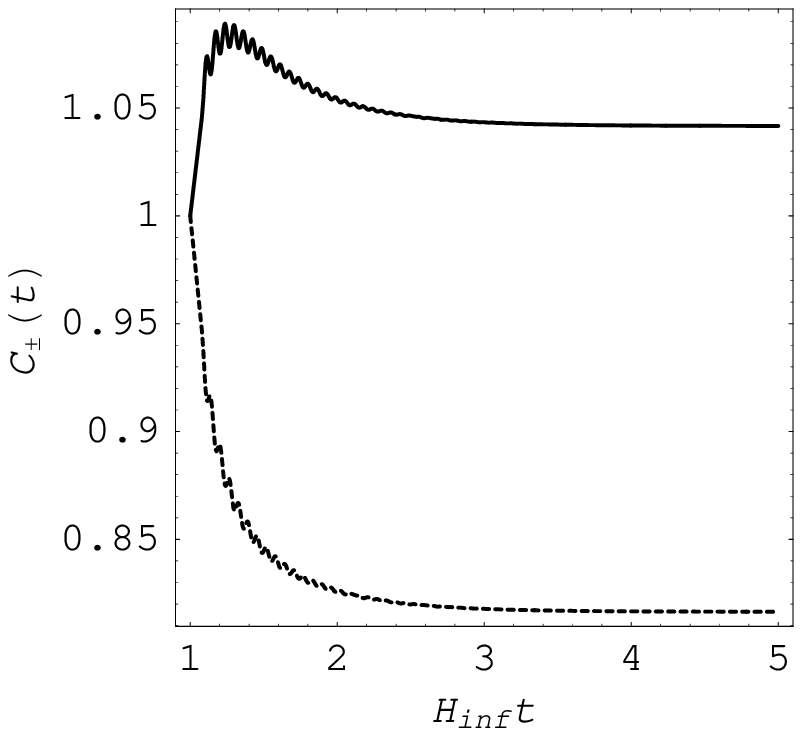}
\caption{
Numerical results of the evolution of $C_{\pm}(t)$ 
for $H_{\mathrm{inf}} = 10^{10}$GeV, $m = 10^{12}$GeV, 
$\lambda = - 135$, 
$\tilde{\lambda}=1.0$
$w=1/(75)$, $\phi_1 = 10^{12}$GeV, $M = 10^{12}$GeV, and 
$g_{\mathrm{ps}} = 1.0$ (the case (iii) in Table \ref{table:1}).  
Here 
we have used the evolution of $\phi$ in Eq.\ (\ref{eq:3.8}).  
The solid curves represents $C_+(t)$ and the dotted curves 
represents $C_-(t)$.   
}
\end{center}
\end{figure}

\end{document}